# QoS Provisioning Using Hybrid FSO-RF Based Hierarchical Model for Wireless Multimedia Sensor Networks

Saad Ahmad Khan , Sheheryar Ali Arshad

Department of Electrical Engineering,
University Of Engineering & Technology, Lahore
Pakistan, 54890

Email: saad.ahmad@uet.edu.pk; s.ali@uet.edu.pk

*Abstract-* **Our objective is to provide guaranteed packet delivery service in time constrained sensor networks. The wireless network is a highly variable environment, where available link bandwidth may vary with network load. Since multimedia applications require higher bandwidth so we use FSO links for their transmission. The main advantage of FSO links is that they offer higher bandwidth and security, while RF links offer more reliability. The routing in this multi-tier network is based on directional geographic routing protocol, in which sensors route their data via multi-hop paths, to a powerful base station, through a cluster head. Some modifications have also been incorporated in the MAC layer to improve the QoS of such systems.**

## I. INTRODUCTION

RECENT advancement in field of sensor networks show that there has been increased interest in the development multimedia sensor network which consists of sensor nodes that can communicate via free space optics (FSO) or RF. A wireless multimedia sensor network typically consists of two types of sensor nodes. One of these acts as data sensing nodes with sensors like acoustic sensors or seismic sensors etc. The other nodes are the video sensor nodes which capture videos of event of interest.

Multimedia contents, especially video streams, require transmission bandwidth that is orders of magnitude higher than that supported by current off-the-shelf sensors. Hence, high data rate and low-power, consumption-transmission techniques must be leveraged. In this respect, free space optics seems particularly promising for multimedia applications.

FSO refers to the transmission of modulated visible or infrared (IR) beams through the atmosphere to obtain broadband communications over distances of several kilometers. The main limitation of FSO is the requirement that a direct line-of-sight path exist between a sender and a receiver. However FSO networks offer several unique advantages over RF networks. These include the fact that FSO avoids interference with existing RF communications infrastructure [1], is cheaply deployed since there is no government licensing of scarce spectrum required, is not susceptible to "jamming" attacks, and provides a convenient bridge between the sensor network and the nearest optical fiber. In addition, "well-designed" FSO systems are eye safe, consumes less power and yields smaller sized nodes because a simple baseband analog and digital circuitry is required, in contrast to RF communication. More importantly, FSO networks enable high bandwidth burst traffic which makes it possible to support multimedia sensor networks [1].

| *Class* | *Application* | *Bandwidth (b/s)* | *Delay bound (ms)* | *Loss Rate* |
|---|---|---|---|---|
| **Non-real time variable bit rate** | Digital Video | 1M – 10M | Large | $10^{-6}$ |
| **Available Bit Rate** | Web Browsing | 1M - 10M | Large | $10^{-8}$ |
| **Unspecified Bit Rate** | File Transfer | 1M - 10M | Large | $10^{-8}$ |
| **Constant Bit Rate** | Voice | 32 k – 2M | 30-60 | $10^{-2}$ |
| **Real time Variable Bit Rate** | Video Conference | 128k - 6M | 40-90 | $10^{-3}$ |

Table 1 Typical QoS requirements for several service classes

## II. RELATED WORK

Inherently a multi-path protocol with QoS measurements and a good fit for routing of multimedia streams in WSN. Multi-flow Real-time Transport Protocol (MRTP) [2] is suited for real-time streaming of multimedia content by splitting packets over different flows. However, MRTP does not specifically address energy efficiency considerations in WMSNs. In [3], a wakeup scheme is proposed to balance the energy and delay constraints.

In [4], the interesting feature of the proposed protocol is to establish multiple paths (optimal and suboptimal) with different energy metrics and assigned probabilities. In [5], a Multi-Path and Multi-SPEED routing protocol is proposed for WSN to provide QoS differentiation in timeliness and reliability.

In [6], an application admission control algorithm is proposed whose objective is to maximize the network lifetime subject to bandwidth and reliability constraints of the application. An application admission control method is proposed in [7], which determines admissions based on the added energy load and application rewards. While these approaches address application level QoS considerations, they fail to consider multiple QoS requirements (e.g., delay, reliability, and energy consumption) simultaneously, as required in WMSNs.

The use of image sensors is explored in [8], in which visual information is used to gather topology information that is then leveraged to develop efficient geographic routing schemes. A similar scenario is considered in [9] where imaging data for sensor networks results in QoS considerations for routing

Recent studies have considered the effect of unidirectional links [10], and report that as many as 5% to 10% of links in wireless ad hoc networks are uni-directional [11] due to various factors. Routing protocols such as DSDV and AODV which use a reverse path technique implicitly ignore such unidirectional links, and are therefore not relevant in this scenario. Other protocols such as DSR [10],
ZRP [12] or SRL [13] have been designed or modified to accommodate unidirectionality, by detecting unidirectional links, and then providing a bi-directional abstraction for such links [14], [15], [16], [17]. The simplest and most efficient solution proposed for dealing with unidirectionality is Tunneling [18], in which bi-directionality is emulated for a uni-directional link by using bi-directional links on a reverse backchannel to establish the tunnel.

Tunneling also prevents implosion of acknowledgement packets and looping by simply repressing link layer acknowledgments for tunneled packets received on a unidirectional link. Tunneling however works well in a mostly bi-directional network with few unidirectional links [10].

Our contribution in two manifold. We've given a novel routing algorithm and also introduced a novel approach to improve QoS at Network and MAC layer. In Section III we propose the Hybrid based RF-FSO System which includes the routing model, a novel routing approach to send the aggregated data via FSO links to sink and a suitable Medium Access Control Layer Protocol to improve the Quality of Service

## III. PROPOSED HYBRID FSO/RF BASED SYSTEM

The key observation to our hybrid architecture is that in wired networks, the delay is independent of the physical distance between the source and sink, but in case of multi-hop wireless sensor networks, the end-to-end delay depends on not only single hop delay, but also on the distance a packet travels.
In view of this, the key design goal of such architecture is to support a soft real-time communication service with a desired delivery *speed* across the sensor network, using FSO links for high bandwidth applications and RF links for initiating routing paths and low bandwidth applications.

### NETWORK MODEL
In Figure 1, the base station is elevated over the RF/FSO WSN deployment area. There are two types of channels in the RF/FSO WSN:
1) RF peer-to-peer channels

2) Narrow line of sight (LOS) FSO channels that connect the cluster heads to the base station.
The FSO link is achieved by using a passive modulating optical retro-reflector mounted on each node. The base station steers a narrow optical beam to interrogate all the FSO nodes in the field. Nodes that are too far from the base station, or which do not have Line Of Sight to the base station, communicate with the base station through RF peer-to-peer multi-hop links. Due to this some of the nodes in the network act as cluster heads.

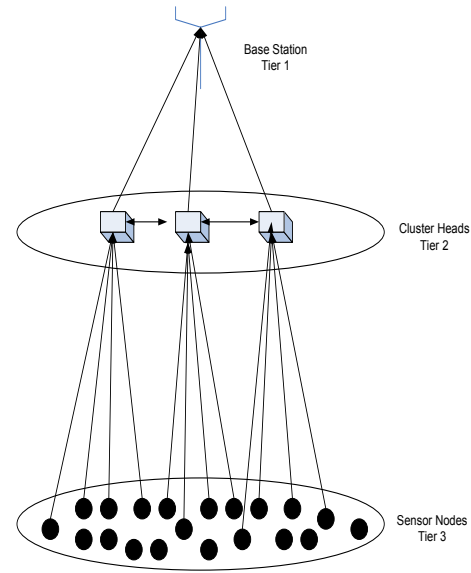

Figure 1- Multiple Tier Network Structure for Hybrid RF-FSO

In such RF/FSO based wireless sensor network, none of the nodes communicate directly to the sink (base station) using RF links.

### ROUTING PROTOCOL DESIGN

We consider a geographic WMSN which consist of finite set of sensor nodes which can be represented as N={ $n_1,n_2,...,n_N$} whereas finite set of links between them are L={$l_1,l_2,...,l_N$}. The location of the base station and the sensor nodes is fixed that can be obtained through GPS. The nodes that are basically at 1-hop distance from the Base Station are Cluster heads that uses FSO link for their communication. The cluster heads are equipped with RF/optical trans-receiver (consisting of Infrared/semiconductor laser and photo-detectors). Each Cluster Head $S_x$ has a position ($x_a$, $y_a$) and directional orientation $\theta_x$, and can orient its transmitting laser to cover a contiguous scanning area, given as

$$\alpha/2 + \theta_x \leq \phi_x \leq +\alpha/2 + \theta_x. \qquad (1)$$

Following the model as depicted in Figure 2, each cluster head $S_x$ can send data over oriented sector $\phi_x$ of $\alpha$ degrees, for a fixed angle $0 < \alpha < 2\pi$. The receiving photo-detector is omni-directional and can thus receive data from any direction, so that the sensing region of the cluster head is not only limited to its communication sector. For another cluster head $S_y$ to receive data from $S_x$ two conditions must be met:

1) The distance between them should be lesser than the communication range R(n) of the sender cluster head, i.e., $D(S_x, S_y) \leq R(N)$.

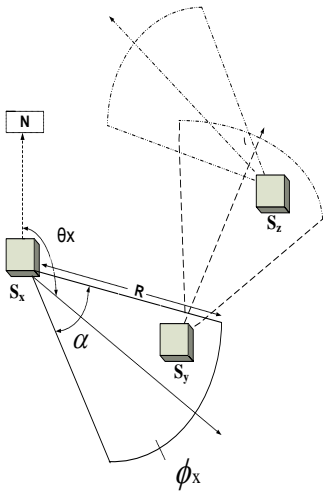

Figure 2 Cluster head $S_y$ falls into communication sector of $S_x$

**2)** The receiving cluster head must be within the $\phi_x$ of the sender cluster head, i.e., $(x_b, y_b) \in \phi_x$, where $(x_b, y_b)$ defines the location of the receiver cluster head. For this setup, $S_x$ may directly talk to $S_y$; however, $S_y$ can only talk to $S_x$ via reverse route, with other cluster heads in the network acting as simple routers [1].

Let us suppose that cluster head $S_x$ is initiating next hop selection to find routing routes. The coordinates of a $S_x$ $(x_a, y_a)$ is piggybacked in the message broadcast by $S_x$. Thus a neighbor node $S_y$ know the position of tits upstream node $S_x$, its own position, i.e., $S_y$ $(x_b, y_b)$ and the sink's location. Further we assume that each cluster head knows about its neighborhood (any algorithm can be used to find out the neighborhood of a cluster head, e.g., Neighborhood discovery algorithm (NDA) as proposed by [1]). Since cluster heads are equipped with hybrid RF/FSO links, so the first phase in our design is to discover multiple routing paths from source cluster heads to sink cluster head or base station. To establish a path, a probe (COMB) message is broadcast to every 1-hop neighbor initially by the source for route discovery. The selected next hop CH will continue to broadcast COMB message to find their next hop, and so forth until sink CH is reached.

The information that is contained in a COMB message is shown below

| Fixed Attributes | | |
|---|---|---|
| SourceID | SinkID | |
| DeviationAngle | SrcToSinkHopCout | |
| Variable Attributes | | |
| HopCount | PreviousHop | Position |

Figure 3- Packet format of COMB

The COMB message is identified by the SourceID, SinkID. DeviationAngle (denoted by α) specifies the unique direction where a path will traverse during path establishment. The fixed attributes in a COMB are set by the source and not changed while the COMB is propagated across the network. On the other hand, when an intermediate CH broadcasts a COMB, it will change the variable attributes in the message. Hop-Count is the hop count from the source to the current CH.

PreviousHop is the identifier of the current CH. Position denotes the absolute coordinates of the current CH.

A cluster head receiving a COMB will calculate its distance from the destination/sink CH. If the distance between the cluster head that received the COMB message and the sink CH is lesser than the distance between the source CH to the sink CH and the CH that received the PROBE message lies inside the communication sector of source CH, then that CH will become the next hop CH. The same procedure will be repeated for all other CH's and multiple delivery guaranteed routing paths from Source CH to sink CH will be discovered that will use FSO link for multimedia transmissions.

The next phase is to find the best possible routing path from already explored multiple paths. In order to do so, we assume a Reference Line directly connecting source CH and the destination CH.

For each and every path, we calculate the distance between every CH that comes along that path and the Reference Line and then take its average.

$$Dpath \cong \frac{\sum_{i=1}^{N} d_i}{\text{Hopcount-1}}$$

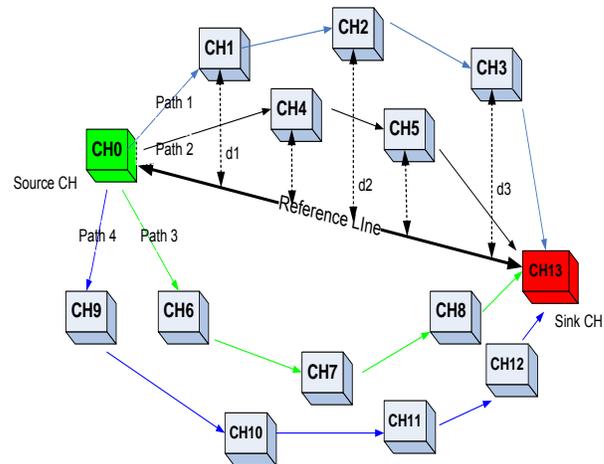

Figure 4 – Mulitpath Establishment Phase

where, $d_i$ for Path1 is the sum of d1, d2 and d3 which are the distances of $CH_1$, $CH_2$ and $CH_3$ from the Reference Line.

Similarly, we will find $D_{path2}...,D_{pathM}$ where M are the total number of routing paths explored from source CH to sink CH. We will select that path for routing multimedia application which has got smallest value of $D_{path}$. Once the best path has been explored, then we can use FSO links and corner cube retro reflector CCR for bandwidth hungry applications.

## IV. MAC QoS PROVISIONING

Research efforts to provide MAC layer QoS can be classified mainly into (i) channel access policies, (ii) scheduling and buffer management, and (iii) error control.

**(i) Channel Access Policy**

Motivated by the power-aware protocol [20], we define Cluster-head Reservation based MAC scheme (CRB-MAC).
We make the following assumptions:
i. All nodes are synchronized to a global time, according to some synchronization scheme and time is sliced in frames of fixed length.
ii. Loss of packets can only occur due to collisions or missed deadlines.
iii. Admission control has been performed already, so that the offered load can be sustained by the network.
iv. The network is stationary, i.e., there is no mobile node.

Let $Y(i)$ be a node in the network such that total number of nodes $Y_{total} = \sum_{i=1}^{N} Y_i$. The range $R(x)$ of a cluster-head 'x' contains a set of cluster-heads within its RF/FSO range.
$R(x) = \{h \mid h$ is in transmission range of cluster-head$\}$
There is a set $R(x, y)$ which contains all set of cluster heads in common range of two cluster-heads
$R(x, y) = R(x) \cap R(y)$

We assume that the geographic area of the network is divided into rectangular *grids* of equal size. Each gird of size $C_a \times C_b$ covers at least one cluster head. Each cluster head has its geographic location (found via GPS) which is mapped on one-to-one basis to the grid location.

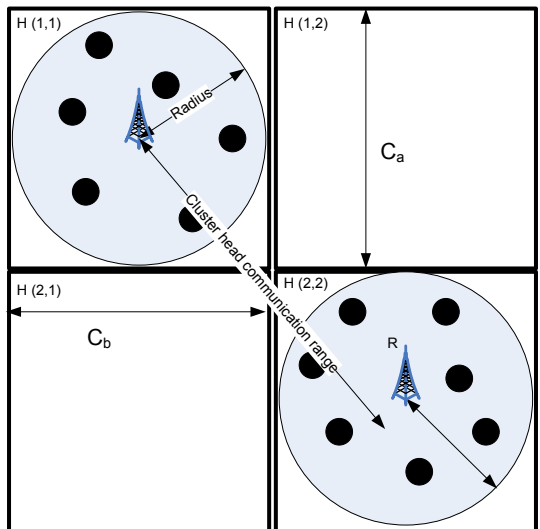

Figure 5 - Cluster Head Grid

Each frame is divided, into the Reservation Period (RP), during which nodes compete to establish new reservations or cancel reservations, and the Contention-Free Period (CFP), during which they send data packets without contention during the reserved transmission windows and sleep when they do not have a packet to send or receive.

The Reservation Period is divided in "Reservation Period slots (RP-slots)", each having a fixed length (which is enough for three reservation messages to be transmitted).

The Contention-Free period is divided in a Contention-Free (CF) slots. Each CF-slot has fixed length, long enough for a transmission of a data packet and an acknowledgment.

Each station δ keeps the reservation information in a Reservation Table (RT), which keeps track of the IDs of the nodes (within range) that are transmitting or receiving data during each Contention Free slot. When a node joins the network it has to stay awake for some period to hear the ongoing transmissions and to update its reservation table.

*Reservation Period*

During the Reservation Period, two types of "reservation procedures" can take place, i.e., the *Connection Establishment* procedure and the *Connection Cancelation* procedure.

A station that needs to initiate a Connection Establishment or Connection Cancelation can do so in the pre-specified Reservation Slot for its grid.

A host in a grid $H(x, y)$ can initiate a reservation procedure only during the reservation slot '$r$' such that $r = 3x+2y+5$. This ensures that only one reservation in $H$ a rectangular area of **3x3** grids can take place in one reservation slot.

| T(1, 1) | T(1, 2) | T(1, 3) |
| --- | --- | --- |
| T(2, 1) | T(2, 2) | T(2, 3) |
| T(3, 1) | T(3, 2) | T(3, 3) |

Figure 6- Reservation Slot using Grid

When a station needs to make a connection establishment or cancellation, it senses the channel to determine if another station of the same grid is transmitting. The station proceeds with a transmission if the medium is determined to be idle for a specified interval. If the medium is found to be busy, or the first message is not sent successfully, by a sender then the exponential back-off algorithm is followed and the station chooses a subsequent frame to re-initiate its request.

**The *Connection Establishment*** procedure takes place for a real-time station every time a new real time session begins. Datagram (non-real-time) traffic is not sensitive to delay, thus nodes may buffer the packets up to a "burst length" N and then make a request for sending the whole burst. The reservation establishment involves the exchange of three control messages:

(a) A Connection Request $CR(x, y)$ is sent by a node $x$ to a neighbor $y$ for which it has real-time or non-real-time packets.

| Packet length | Free Slots | Deadline for the real-time data packet |
| --- | --- | --- |

Figure 7(a) Real time data packet format

| Packet length | Free Slots | Number of buffered packets of non-real time data to be sent |
|---|---|---|

Figure 7(b) Non-real time data packet format

(b) A Connection Acknowledgment *CA(y, x)* is sent by a node *y* to a neighbor *x*. The CR from x contains the information of free slots of *x*. Node y scans for all of its free slots and compares it with the free slots of *x*.
Then, it schedules the requested packet in a common free slot in its Reservation Table. Then, the receiver indicates in the CA which reserved slot(s) the sender should use.

$RT(i) = \{F_s | F_s \in \{X_{Fs} \cap Y_{Fs}\}\}$

(c) A Slot Rese *SRB (x, y)* is sent by a node *x* to all other nodes and includes the reserved slots that *x* has reserved. Thus all the nodes in neighborhood become also aware of the reservation.

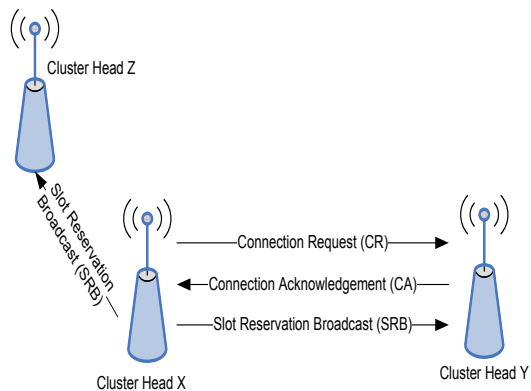

Figure 8- Connection Establishment between Node X and Node Y

**The *Connection Cancelation*** is invoked when a sender has no more packets to send during its real-time session. Two messages are involved in the Reservation Cancelation:
(a) The Connection Cancel *CC(x, y)* sent by a node *x* to node *y*
(b) The Connection Cancel Acknowledgment *CC-ACK(y, x)*, sent by node *y* to *x*.

*Contention-Free Period*
During the CFP, the stations wake up in the predetermined transmission slots according to their Reservation Table, to send or receive data packets, and sleep the rest of the period. In each slot, the sender sends a data packet, with size specified by the sender and receives an acknowledgment (ACK) sent by the receiver. If a node does not have any data to send or receive during a Contention Free slot, then it switches off. Once the reservation for a real time session is established, it is kept until an explicit reservation cancelation is performed as described above. The sender will use the reserved slot to send its data packets until the session is over. Reservations for datagram traffic are valid only for the current frame and the hosts clear their Reservation Table for those slots that have non-real-time transmissions, after the CFP is over. Thus no explicit cancelation is needed in case of datagram reservations.

ii) Scheduling and Buffer Management

The foundation of a proper QoS provisioning lies in the appropriate service model that describes a set of offered services. Existing QoS aware networks, like ATM, InteServ, and DiffServ, designed their service models based on QoS requirements of applications in the corresponding network infrastructure. We have laid the foundation of our service model for hybrid fso-rf based multimedia network.
First the multimedia data is classified into real time and non real time data according to the delay requirements. For achieving desired QoS for a real time application it is usually mandatory to maintain a specific bandwidth during real time transmission time.
We categorize applications to the following classes.

• *Bonded service* represents real-time applications that require absolute continuity, i.e., the transmission of data cannot be disconnected between the session.
• *Semi-bonded service* represents real-time applications that can tolerate a reasonably low disconnection in-between the transmission.

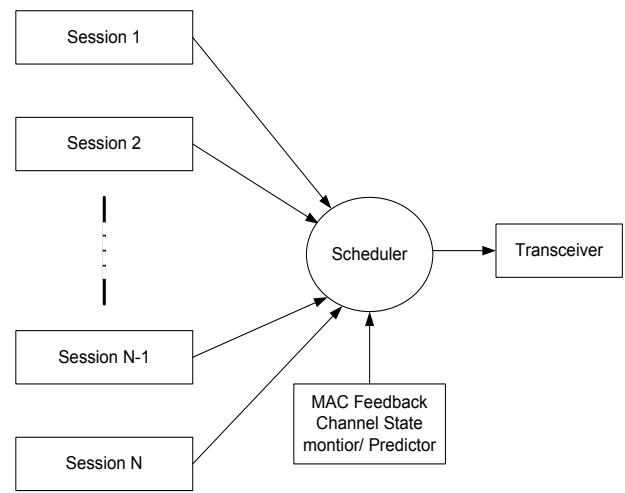

Figure 9- Wireless Multimedia Scheduler

### V. MAC QoS PROVISIONING PROOF

The defined changes in MAC layer provide better QoS under the assumptions as taken before in the paper. The hidden node problem causes the collision and certain critical information is lost. If connection establishment information is lost then the reservation tables can become inconsistent and collisions in data packets may occur or Contention Free slots maybe wasted. When a CR packet is collided, no reservation information is lost. When CA or SRB packets are lost, conflicting reservations can happen, which may result in data packets collisions. When Connection Cancellation or CC-ACK packets are lost, then reservation cancelation information maybe lost and the slots may not be able to be reserved for other hosts, thus data slots remain unused.
We assume that a node initiates a reservation procedure with node that involves CF-slot. To prove MaC QoS provisioning we use the following lemma.

*Lemma 1:* A node k in the network can cause a reservation message to be missed by a node in during Connection

Establishment phase, if and only if, it is couple of hops away, i.e., 1,2,3 or 4 hops-way from the sender

*Lemma 2:* All Connection Reservation Messages are received successfully by the nodes during the time that any reservation procedure is taking place, if and only if, any node initiates a reservation procedure at reserved slot.

*Lemma 3:* The protocol ensures that all nodes insuccessfully update their reservation tables whenever connection establishment or connection cancellation procedure take place.
.
## VI. CONCLUSION AND FUTURE WORK

We have presented a hybrid FSO/RF based model for wireless multimedia sensor network. We have proposed a new routing protocol for such networks to provide energy efficient real time communication. As future work we plan to simulate our protocol and compare it with similar reservation based protocol. It is expected that our protocol consumes less energy for routing multimedia data with minimum delay. At MAC layer we use a fully distributed reservation scheme which is able to provide bandwidth guarantees and energy conservation using geographic information.